\documentclass{aa}
\usepackage[dvips]{graphics}
\usepackage{latexsym}

\def\gtsima{$\; \buildrel > \over \sim \;$}
\def\ltsima{$\; \buildrel < \over \sim \;$}
\def\prosima{$\; \buildrel \propto \over \sim \;$}
\def\gsim{\lower.5ex\hbox{\gtsima}}
\def\lsim{\lower.5ex\hbox{\ltsima}}
\def\simgt{\lower.5ex\hbox{\gtsima}}
\def\simlt{\lower.5ex\hbox{\ltsima}}
\def\simpr{\lower.5ex\hbox{\prosima}}

\def\h1{$h^{-1}$}

\def\beq{\begin{equation}}
\def\eeq{\end{equation}}

\begin{document}


\title{
The Spatial Clustering of Distant, \boldmath{$z\sim1$}, Early-Type Galaxies
}

\author{
	E. Daddi \inst{1}
	\and
	T. Broadhurst \inst{2}
	\and
	G. Zamorani \inst{3}
	\and
	A. Cimatti \inst{4}
	\and
	H. R\"ottgering \inst{5}
	\and
	A. Renzini \inst{2}
}
\institute{ 
	Universit\`a degli Studi di Firenze, Dipartimento di Astronomia e Scienza dello Spazio, Largo E. Fermi 5, 50125 Firenze, Italy
	\and
	European Southern Observatory, 85748 Garching, Germany
 	\and 
        Osservatorio Astronomico di Bologna, Via Ranzani 1, 40127 Bologna, Italy
	\and
	Osservatorio Astrofisico di Arcetri, Largo E. Fermi 5, 50125 Firenze, Italy
	\and
	Sterrewacht Leiden, Postbus 9513, 2300 RA Leiden, The Netherlands
}

\offprints{Emanuele Daddi, \email{edaddi@arcetri.astro.it}}
\date{Received 18 April 2001; accepted 18 July 2001}

\abstract{ We examine the spatial clustering of extremely red objects
(EROs) found in a relatively large survey of 700 arcmin$^2$,
containing 400 galaxies with $R-Ks>5$ to $Ks=19.2$. A comoving correlation length
$r_0=12\pm3$ \h1 Mpc is derived, under the 
assumption that the selection function is described by a passively
evolving early-type galaxy population, with an effective redshift
of $z\sim 1.2$. This correlation length
is very similar to that of local $L^{*}$ elliptical
galaxies implying, at face value, no significant clustering evolution
in comoving coordinates
of early-type galaxies to the limiting depth of our sample, $z\sim1.5$. A rapidly
evolving clustering bias can be designed to reproduce a null result;
however, our data do not show the corresponding strong reduction in the
average population density expected for consistency with underlying
growth of the mass-function. 
We discuss our data in the context of recent ideas regarding bias evolution.
\\ The uncertainty we quote on $r_0$
accounts for the spikey redshift distribution expected along
relatively narrow sightlines, which we quantify with detailed
simulations. This is an improvement over the standard use of Limber's
equation which, because of its implicit assumption of a smooth selection function,
underestimates the true noise by a factor of
$\approx 3$ for the parameters of our survey. We propose a
general recipe for the analysis of angular clustering, suggesting that
any measurement of the angular clustering amplitude, A, has an
intrinsic additional uncertainty of $\sigma_A/A=\sqrt{AC}$, where $AC$
is the appropriate integral constraint.  
\keywords{Cosmology: large-scale structure of Universe; Galaxies: evolution; Galaxies:
elliptical and lenticular, cD; Galaxies: formation; Galaxies: fundamental parameters}
} 
\titlerunning{Spatial Clustering of $z\sim1$ Early-Type Galaxies}
\authorrunning{E. Daddi et al.}  
\maketitle

\section{Introduction}

The evolution of the galaxy two-point correlation function provides
important insights into the nature of galaxy formation and evolution
(Peebles 1980, Efstathiou et al. 1991).  
The shape and normalization of this function depends on both
the cosmic growth of mass structures and on the details of how
galaxies trace mass at different epochs - the bias
evolution. Measurements of the clustering evolution of galaxies of
different luminosities and morphological types help constraining how,
when and where they were formed.

In the last few years, a number of investigations of the spatial clustering of
distant galaxies has been carried out. A marked decline in the amplitude
of the spatial correlation function has been reported to 
$z\sim1$, for magnitude selected samples of field galaxies (Le
Fevre et al. 1996, Carlberg et al. 1997, Hogg et al. 2000), consistent with a stable
clustering scenario.  This decline seems to reverse 
towards high--$z$, given the strong clustering reported
for Lyman-Break Galaxies (LBG's hereafter) at $z=3$ (Giavalisco et
al. 1998, Adelberger et al. 1998).

Recently, Daddi et al. 2000 (D2000 hereafter) detected a large angular
clustering signal from extremely red, $R-Ks>5$, galaxies (EROs)
obtained from a $K$-selected survey  over 700 arcmin$^2$. 
The angular clustering of
EROs was found to be an order of magnitude larger than the full
$K$-magnitude selected galaxies, and an increase of the clustering
signal was detected with increasing $Ks$ luminosity and
increasing $R-Ks$ color (D2000).  A similarly large angular clustering
amplitude for EROs has been reported by McCarthy et al. (2000) in an imaging
survey of red galaxies detected in an $H$--selected sample.

From the selection
criteria it is known that EROs can be high-redshift passively evolving
ellipticals or dusty starbursts, and examples of both classes exist
(see e.g. D2000 for more details). It is becoming clear however, that
the bulk of the ERO population is probably dominated by the former class: Broadhurst
\& Bouwens (2000), Moriondo, Cimatti \& Daddi (2000) and Stiavelli \&
Treu (2000) have concluded from independent datasets that most ($\sim 70$\%)
such objects have De Vaucouleurs profiles, with only about 15\% of them
displaying irregular or disturbed morphologies, expected for dusty
starburst systems (the remaining 15\% has a disk-like exponential profile). 
The existing spectroscopic results for single
objects or for small samples of EROs support this conclusion (Spinrad
et al. 1997; Soifer et al. 1999; Cimatti et al. 1999, Liu et
al. 2000). Spectroscopy of flux-limited samples of $K$ selected galaxies
generally suffer from incompleteness of the optically reddest galaxies
but the weight of evidence is that most of the reddest objects have spectra
consistent with early-type galaxies, up to the effective spectroscopic
limit of $z\sim1.3$ for absorption-line work (Cohen et al. 1999;
Eisenhardt et al. 1998; Cimatti 2001). These
surveys also broadly show that the EROs of known redshift
are in the range $0.8\simlt z\simlt1.5$,
consistent with expectations for passively evolving ellipticals
based on an extrapolation of the local luminosity function with
passive evolution (Sect. \ref{sec:dNdz}).

Even if a few EROs have been identified as counterparts of SCUBA
sources (Smail et al.  1999; Gear et al. 2000), SCUBA observations
of complete ERO samples show no frequent detections (Mohan et
al. 2001, in preparation) reinforcing the idea that dusty HR10-like 
objects (Cimatti et al. 1998, Dey et al. 1999, Andreani et al. 2000) are rare among EROs.
It is therefore reasonable to conclude that with $K\simlt19$ EROs we
are observing the clustering signal of predominantly distant
early-type galaxies.

Here we take the angular correlation function measurements of the EROs
and a plausible estimate of their selection function to
derive their spatial clustering amplitude. The 3D correlation
length of EROs should thus produce constraints on the evolution of the
clustering of early-type galaxies, a population of objects which
is likely to have formed in the highest amplitude perturbations 
and to be positively biased
with respect to the general galaxy population and to the overall
distribution of mass. The evolution of the correlation amplitude of
early-type galaxies is an observable which is independent of
measurements of the evolution of their comoving number density,
providing therefore a complementary mean to examine the rate of
evolution. Constraints on the density evolution of EROs have been
sought previously in response to predictions of CDM models (Baugh et
al. 1996, Kauffmann 1996). 
Early work based on small fields claimed observational
evidence for a sharp decline in space density of early type galaxies,
while more recent estimates based on deeper and more
complete samples are consistent with a constant comoving density of this population
up to at least $z\sim1.3$, and imply a typical formation
redshift of the stars in these galaxies 
not less than $z_f\sim2.5$ assuming passive evolution
(see Daddi, Cimatti \& Renzini 2000 for a complete discussion).  Hence
it is very important to obtain information regarding the clustering
evolution to independently address this important question.  
Here we analyze together both these questions of clustering and density
evolution of EROs, with the largest complete sample of relevant data.

In Sect. \ref{sec:Limber} we describe the standard Limber's equation
formalism. Sect. \ref{sec:dNdz} derives order-of-magnitude
results on the correlation length $r_0$ that confirm the interpretation of EROs as
high--$z$ early type galaxies and justify our assumed
redshift distribution. Sect.  \ref{sec:Alimb} present the
$r_0$ estimates based on Limber's equation. In Sect. \ref{sec:simul} we
discuss the limitation of the standard approach and we use
numerical simulations to
find the definitive constraint on the correlation length, that differs
significantly from the Limber's equation results. We discuss the
general implications of our findings for the clustering analysis on
small areas. We than compare in Sect. \ref{sec:evol} our estimates of
the correlation length of distant ellipticals to the measurements in
the local universe and to theoretical models predictions. Our
conclusions are presented in Sect. \ref{sec:conclu}.

All the scales quoted in the paper are given in comoving units.  Three
cosmological models have been considered: a $\Lambda$-flat universe
($\Omega_m=0.3$, $\Omega_\Lambda=0.7$, $h=0.7$), an open universe
($\Omega_m=0.3$, $\Omega_\Lambda=0$, $h=0.7$) and an $\Omega$-flat universe
($\Omega_m=1$, $\Omega_\Lambda=0$, $h=0.5$).  $H_0=100 h$ km s$^{-1}$ Mpc
$^{-1}$.

\section{The Limber's equation and the spatial correlation length}
\label{sec:Limber}

The angular two point correlation function $w(\theta)$ is related to
its real space analogous $\xi(r)$ by Limber's equation (Peebles
1980).  In the case of small angles ($\theta\ll1$), if both $w$ and
$\xi$ have power law shapes, writing $\xi(z)=(r/r_0(z))^{-\gamma}$ (with $r_0(z)$ being the
comoving correlation length at redshift $z$, and $r$ the comoving distance), the Limber's equation
becomes: 
\beq w(\theta) = \sqrt[]{\pi}
\frac{\Gamma((\gamma-1)/2)}{\Gamma(\gamma/2)} \frac{\int
g(z)(dN/dz)^2r_0(z)^\gamma dz}{[\int (dN/dz) dz]^2} \theta^{1-\gamma}
\label{eq:Limber}
\eeq
where $dN/dz$ is the redshift selection function of the sample, which in the limit of a 
large number of objects coincides with the observed redshift distribution.
The function g(z) depends only on the cosmology:
\beq
g(z) = (dx/dz)^{-1}x^{1-\gamma}F(x)
\label{eq:gz}
\eeq
where $x$ and $F(x)$ are defined with the metric:
\begin{displaymath}
ds^2=c^2dt^2-a^2[dx^2/F(x)^2 + x^2(d\theta^2+sin^2\theta d\phi^2)].
\end{displaymath}

If we define:
\beq
r_{\rm 0,eff}^\gamma = \int g(z)(dN/dz)^2r_0(z)^\gamma dz\  {\LARGE /} \int g(z)(dN/dz)^2
\label{eq:r0}
\eeq
then by using eq. \ref{eq:Limber} and with $w(\theta)=A\theta^{1-\gamma}$ we
have:
\beq 
A = r_{\rm 0,eff}^\gamma B
\label{eq:A}
\eeq
\beq
B=\sqrt[]{\pi} \frac{\Gamma((\gamma-1)/2)}{\Gamma(\gamma/2)} \frac{\int g(z)(dN/dz)^2dz}{[\int (dN/dz) dz]^2}
\eeq

Thus, knowledge of a measured or assumed redshift distribution allows us to relate the
angular clustering amplitudes $A$ to the 3D correlation length $r_{\rm 0,eff}$.
If  $d^2r_0(z)^\gamma/dz^2$ is negligible in the relevant redshift range,
then $r_{\rm 0,eff}=r_0(z_{\rm eff})$ with:
\beq
z_{\rm eff} = \int zW(z)dz
\label{eq:W}
\eeq
\beq
W(z) = g(z)(dN/dz)^2 / \int g(z)(dN/dz)^2dz 
\eeq

This {\it inversion} process provides a weighted estimate of $r_0$
over the probed redshift range.  In the following, we will refer
generically to $r_0$ as $r_{\rm 0,eff}=r_0(z_{\rm eff})$, bearing in
mind, anyway, the effect of eqs. \ref{eq:r0} and \ref{eq:W}.
For consistency with D2000, where the clustering amplitudes were
fitted with $w(\theta)\propto\theta^{-0.8}$, we assume $\gamma=1.8$
for our analysis, consistent with most previous observations (see
e.g. Roche \& Eales 1999). In the following we will quote the values for the amplitude of 
the angular two-point correlation function as measured (or extrapolated) to 1 degree, i.e.
$A\equiv A(1^o)$.

\section{The effects of the selection function}
\label{sec:dNdz}

To derive 3D information from the angular correlation measurements a
selection function for EROs must be supplied. As discussed in the introduction, 
strong evidence exist that the bulk of EROs is made of early type galaxies.

The strong angular clustering of EROs reported by D2000 and McCarthy et al. (2000), 
independently supports this same conclusion. In fact, early-type
galaxies in the local universe are known to prefer high density environments (Dressler 1980)
and to have much larger correlation lengths than late-type galaxies ($r_0\simgt7$--8 versus 
$r_0\simlt5$; Davis \& Geller 1976; Giovanelli, Heynes \& Chincarini 1986; Loveday et
al. 1995) and than dusty starburst (see e.g. Saunders et al. 1992 for IRAS galaxies
that have $r_0\sim3.8$ \h1 Mpc).
At higher ($z\sim1$) redshift even lower correlation lengths 
have been observed for star forming galaxies:
Adelberger et al. (2000) find that {\it balmer break} ($z\sim1$ star forming) galaxies
have $r_0\simlt3$ \h1 Mpc, while the blue starburst selected field galaxies at $0.8<z<1.5$ 
have $1\simlt r_0\simlt 2.5$ \h1 Mpc (Carlberg et al. 1997, Hogg et al. 2000).

\begin{figure}[ht]
\resizebox{\hsize}{!}{\includegraphics{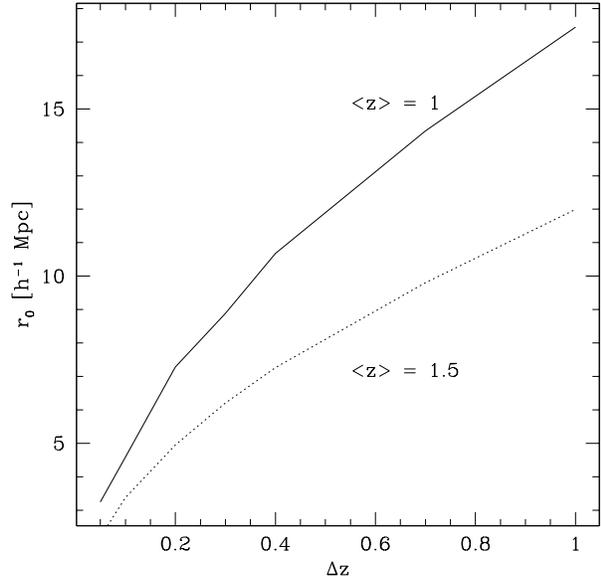}}
\caption{\footnotesize 
The correlation length $r_0$ that reproduces the typical ERO clustering of $A=0.02$ 
(D2000),  as a function of
the width of a top-hat redshift distribution centered at $z=1$ (solid line) 
and $z=1.5$ (dotted line), for the $\Lambda$-flat cosmology.
}
\label{fig:tentative}
\end{figure}

In Fig. \ref{fig:tentative} we calculate
the width of simple top-hat redshift distribution that can
reproduce the ERO observed angular amplitude, as a function of $r_0$.
Requiring that EROs have a correlation length
of $r_0\simlt3$ \h1 Mpc, typically measured for $z\sim1$ star forming
galaxies, would imply a very narrow ERO redshift distribution
of width $\Delta z\approx 0.05$ to reproduce the observed amplitude,
which seems implausible. For any reasonably broad redshift
distribution a much larger $r_0$ is inferred, favoring the larger
$r_0$ known for local early-type galaxies, but irreconcilable with the
small correlation lengths observed for star forming galaxies at both
low and high-$z$.

In the remainder of the paper the selection functions expected for distant early type galaxies
will thus be used.

\subsection{Modeling the selection function}

We adopt models accounting only for passive evolution, appropriate for
early-type galaxies, to estimate the ERO selection function.  
This is well justified for the present analysis
that deals with the clustering of galaxies redder than $R-Ks>5$, 
corresponding to $z\simgt0.8$--$0.9$ for a passively evolving $L^{*}$
elliptical, as several studies have shown that at least up to $z\sim1$
the photometric and density evolution of the elliptical galaxies is consistent
with passive evolution with no number density evolution (Totani \& Yoshii 1997; Franceschini et
al. 1998; Schade et al. 1999; Im et al. 1999; Broadhurst \& Bouwens
2000; Scodeggio \& Silva 2000; Phillipps et al. 2000; Daddi, Cimatti \& Renzini 2000).

\begin{figure}[ht]
\resizebox{\hsize}{!}{\includegraphics{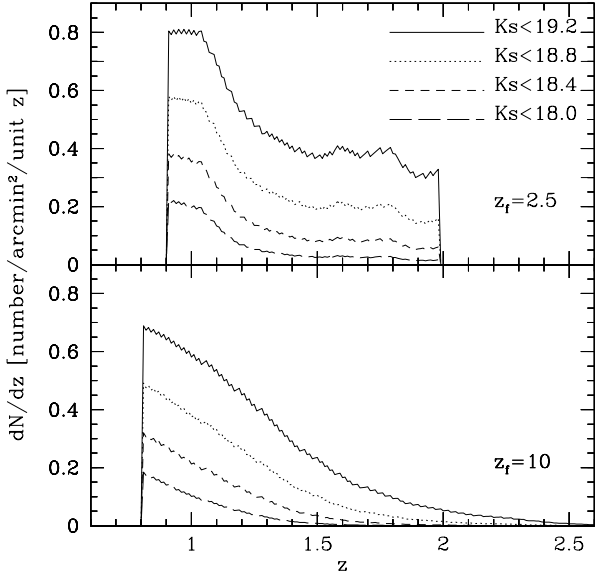}}
\caption{\footnotesize The selection functions of the ellipticals
with $R-Ks>5$ for the passively evolving model described in the text
are shown, for $z_f=2.5$ (top) and $z_f=10$ (bottom), and for various $Ks$
limiting magnitudes, for the $\Lambda$-flat cosmology.
}
\label{fig:dNdz}
\end{figure}

For the passive evolution models adopted here, ellipticals form with a rapid burst
($\tau_{SFR}=0.1$ Gyr).  The Salpeter IMF is assumed, with no dust
reddening, and $Z=Z_\odot$. The Bruzual \& Charlot spectral synthesis
models (1993) in the 1997 version were used, with the Marzke et al.
(1994) pure-ellipticals luminosity function for the normalization at $z=0$.
Daddi, Cimatti \&
Renzini (2000) showed that such models reproduce very well the
ERO number counts in the range $18\simlt K\simlt 20$,
consistently with no appreciable evolution in number density up to $z\sim 1.3$.
In Fig. \ref{fig:dNdz} we show some examples of the redshift selection functions
of $R-Ks>5$ ellipticals, as derived from our models, with
various formation redshift and limiting $Ks$ magnitude.

In conclusion, we have modeled the selection function
accounting only for passive evolution, since empirically
the currently available best data including our own sample,
indicate little (if any) density evolution and SED's and color
trends consistent with pure passive evolution. Some uncertainty
exists at $z>1.5$ where negative density evolution could reduce
the numbers of red objects in a magnitude limited sample, if merging has been important.
However, the selection functions we obtain by setting $z_f$ from 2.5 to 30
bracket a large range of possible models and the difference between
these in terms of the high-z tail shown in Fig. \ref{fig:dNdz} and used in recovering
$r_0$, may be reasonably expected to accommodate
any modest density evolution like that claimed in the models of
Kauffmann et al. (1999), Somerville et al. (2001).
In fact, there are two features of the selection function that mostly 
influence the estimate of $r_0$,
i.e. the width $\Delta z$ and the effective redshift $z_{\rm eff}$ (see Fig. \ref{fig:dNdz}).
Actually, the two limiting cases of $z_f=2.5$ or $z_f=10$ shown in
Fig. \ref{fig:dNdz}, produce only a modest $\sim$10\% variation in the $r_0$ estimate, basically
because the two effects conspire to cancel each other (see Table \ref{tab:chi2}).

\section{Estimating the spatial correlation length with Limber's equation}
\label{sec:Alimb}

\begin{figure}[ht]
\resizebox{\hsize}{!}{\includegraphics{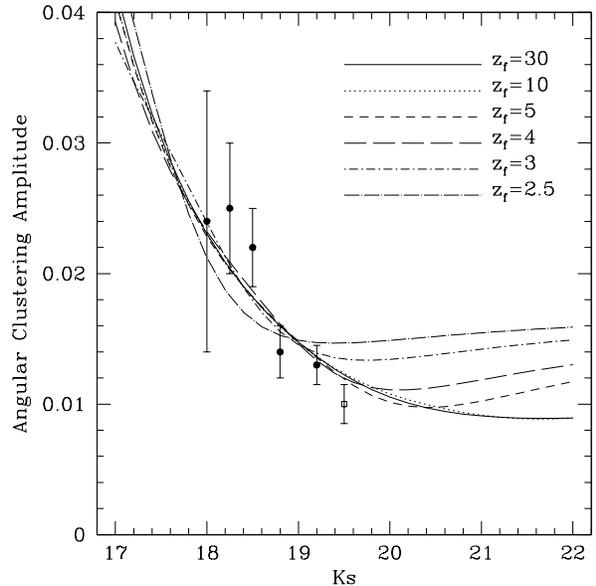}}
\caption{\footnotesize The data are the angular clustering
measurements taken from D2000 (Tab. 5, filled circles). The empty
square is the McCarthy et al. (2000) measurement with $H$ converted to
$Ks$ using $H-Ks=1$.  The passive evolution models predictions are also shown, in
the case of a $\Lambda$-flat cosmology.  The $r_0$ value adopted for
each model is the best fitting one, as shown in Tab. \ref{tab:chi2}.
The general trend shown here is unchanged in different cosmologies.  }
\label{fig:AvsK}
\end{figure}

\begin{figure}[ht]
\resizebox{\hsize}{!}{\includegraphics{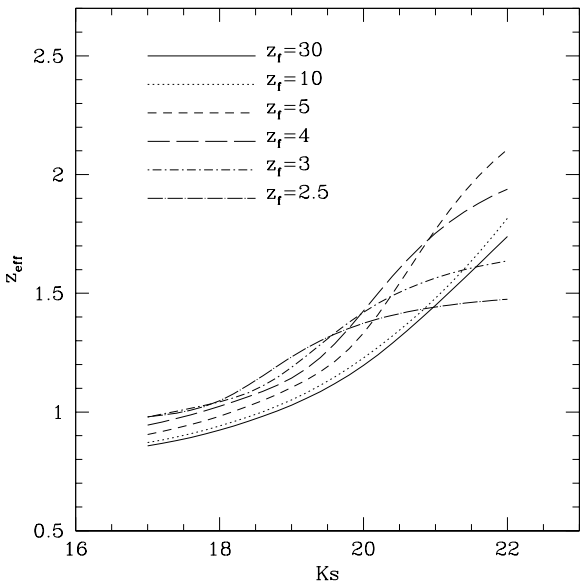}}
\caption{\footnotesize The variation of the effective redshift
$z_{\rm eff}$ (defined in eq. \ref{eq:W}) for samples of passively
evolving ellipticals, selected with $R-Ks>5$, as a function of the $Ks$
limiting magnitude, for the $\Lambda$-flat cosmology.  }
\label{fig:zbar}
\end{figure}

\begin{table}[ht]
\begin{flushleft}
\caption{Real space correlation lengths for EROs, derived through Limber's equation, assuming 
the selection functions expected for the ellipticals in the passive evolution case. The correlation
lengths $r_0$ are expressed in comoving \h1 Mpc, but a proper scaling to $h$ values  different 
from those used in the models would require a recalculation of the selection functions.
}
\protect\label{tab:chi2}
\begin{tabular}{c|cc|cc|cc}
\noalign{\smallskip}
\hline
\noalign{\smallskip}
\multicolumn{1}{c}{} & \multicolumn{2}{c}{$\Lambda$-flat} & \multicolumn{2}{c}{open} & \multicolumn{2}{c}{$\Omega$-flat} \\
\noalign{\smallskip}
\multicolumn{1}{c}{$z_f$} & \multicolumn{1}{c}{$r_0$}  & \multicolumn{1}{c}{$\chi^2_{\rm min}$} &\multicolumn{1}{c}{$r_0$} & \multicolumn{1}{c}{$\chi^2_{\rm min}$} & \multicolumn{1}{c}{$r_0$} & \multicolumn{1}{c}{$\chi^2_{\rm min}$}
\\
\noalign{\smallskip}
\hline
\noalign{\smallskip}
2.5 & 14.1 &  7.1 & 10.6 &  10.5 & 8.3 &  12.0\\ 
3 & 14.6  &  3.6 & 12.5 &  4.2 & 10.3 &  10.3\\ 
4 & 14.3 &  3.1 & 12.3 &  1.9 & 11.7 &  6.5\\ 
5 & 13.9 &  3.7 & 11.6 &  2.3 & 12.3 &  3.8\\ 
10 & 13.3 &  3.7 & 11.3 &  3.7 & 11.8 &  2.2\\ 
30 & 12.9 &  3.6 & 11.4 &  3.4 & 11.2 &  3.0\\ 
\noalign{\smallskip}
\hline
\end{tabular}
\end{flushleft}
\end{table}

In D2000, clustering amplitudes for EROs with $R-Ks>5$ were
estimated at several $Ks$ limiting magnitudes (Fig. \ref{fig:AvsK},
see also Tab.  5 of D2000).  By using the passive evolution $dN/dz$ distribution
appropriate for the different $Ks$ limit, the predictions for the
angular clustering amplitude have been derived at the same $Ks$ limits by
means of eq. \ref{eq:A}, as a function of the $r_0$ values.  The best
estimate for $r_0$ was than obtained from a $\chi^2$ minimization
between all the predicted and observed angular clustering amplitude
(Fig. \ref{fig:AvsK}).

Tab. \ref{tab:chi2} reports the inferred results, for different cosmological models
and redshift of formation.
The Table shows that for any given cosmology the best fit values for $r_0$ are 
not a strong function of the unknown formation redshift. For each cosmology the worst agreement,
as judged from the $\chi^2$ values (see also Fig. \ref{fig:AvsK}) is obtained with the
smallest value of $z_f$. For each entry in the Table we estimate a typical error of the order of
$\Delta r_0\simlt1$ \h1 Mpc, obtained by propagating the measured $\Delta A$ values through eq.
\ref{eq:A}, for the three single most precise $A$ measurements (the error corresponding to 
a $\Delta\chi^2=1$ variation are significantly smaller than that).
Given the small variations of $r_0$ in
Tab. \ref{tab:chi2}, as deduced with different formation redshifts, and their
internal variance, we can conclude that, according to the Limber
equation, EROs have a comoving correlation length of $r_0\sim 13.8\pm1.5$ \h1 Mpc in the
$\Lambda$-flat case, or $r_0\sim 11.5\pm1$ in the open or $\Omega$-flat
case, applying to an effective redshift of $1\simlt z_{\rm eff}\simlt 1.2$.

Fig. \ref{fig:AvsK} shows that the dependence of $A$ on the $Ks$
limiting magnitude is consistently reproduced by the
$dN/dz$ variations, so that the effective correlation length
is not very dependent on the $Ks$ magnitude within the
observed range.  This reflects the expected small variation
of $z_{\rm eff}$ over our samples (from $z_{\rm eff}\sim1$ at
$Ks=18$ to $z_{\rm eff}\sim1.2$ at $Ks=19.2$, see
Fig. \ref{fig:zbar}).  
We stress that the optimal agreement between the 
predicted and observed trend of the ERO angular clustering
versus limiting $Ks$ magnitude is good evidence of the consistency of our modeling of the
selection function based on the passive evolution of the stellar populations.

We also plot in Fig. \ref{fig:AvsK} the angular clustering measurement
by McCarthy et al. (2000), converting their limit of $H<20.5$ by using
the typical color of EROs $H-Ks\sim1$ (Cimatti et al.  1999), and
assuming an error on their measurement similar to our best ones, given
their total area of 1000 arcmin$^2$. Their color selection criterion of $I-H>3$
is roughly consistent with our $R-Ks>5$.  The McCarthy et al. (2000)
point is in good agreement, at least within $\approx 1\sigma$, with
our model's predictions for $z_f>4$ and with the general trend of amplitude versus limiting
magnitude inferred by the D2000 survey.

\section{Estimating the spatial correlation length from numerical simulation}
\label{sec:simul}

Can we trust Limber's equation results, given the known spikey structure
of redshift distributions in pencil-beam surveys? The size of our
field ($22\times 32$ arcmin) corresponds, in fact, to $\sim 17\times 25$  
\h1 Mpc at $z=1$, while in the redshift
direction we sample objects over a range of approximately 1000
\h1 Mpc ($\Lambda$-flat cosmology). The comoving correlation lengths derived from the Limber
equation analysis is therefore of the same order of the projected size of our survey,
suggesting that a very clumpy redshift distributions should be
expected for our sample.  As Limber's equation has formally no
dependence on the extension of the field over which the angular correlations
are measured, we expect that it should apply in the limit in which
the analyzed field of view is large enough so that the observed
redshift distribution approaches the selection function.  But
it is not clear a priori if the Limber's equation should apply
accurately for small fields of view, in which the sampling of
redshifts along the line of sight will vary considerably, dominated by
notable spikes (e.g. Broadhurst et al. 1990, Cohen et al. 1999, Yoshida
et al. 2001).

Since this point has not been investigated previously  we have
embarked upon detailed modeling with realistic simulations. To do this we
created clustered 3D distributions of objects with known input $r_0$
in a suitable comoving volume, we then applied the selection function
$dN/dz$ to these simulated data and projected them on the sky for directly measuring 
the two-point angular correlation function for comparison with the data.

To build up the 3D clustered samples the Soneira \& Peebles (1977,
1978) algorithm was used with a 15 level hierarchy, setting the first
pass radius equal to 50 \h1 Mpc and the step distance of the algorithm to
the proper value to obtain correlations with $\gamma=1.8$. We refer to
the original papers for details and discussions about the algorithm.
By directly measuring the 3D two-point functions we calibrate
the algorithm's parameters in order to reproduce the desired $r_0$
value. Such measurements were done by a simple generalizations to 3D, 
following the approach summarized in D2000. The precision we reach is
better than $r_0/\Delta r_0\simgt50$ over the range $6\simlt r_0\simlt
15$.
\begin{figure*}[ht]
\resizebox{\hsize}{!}{\includegraphics{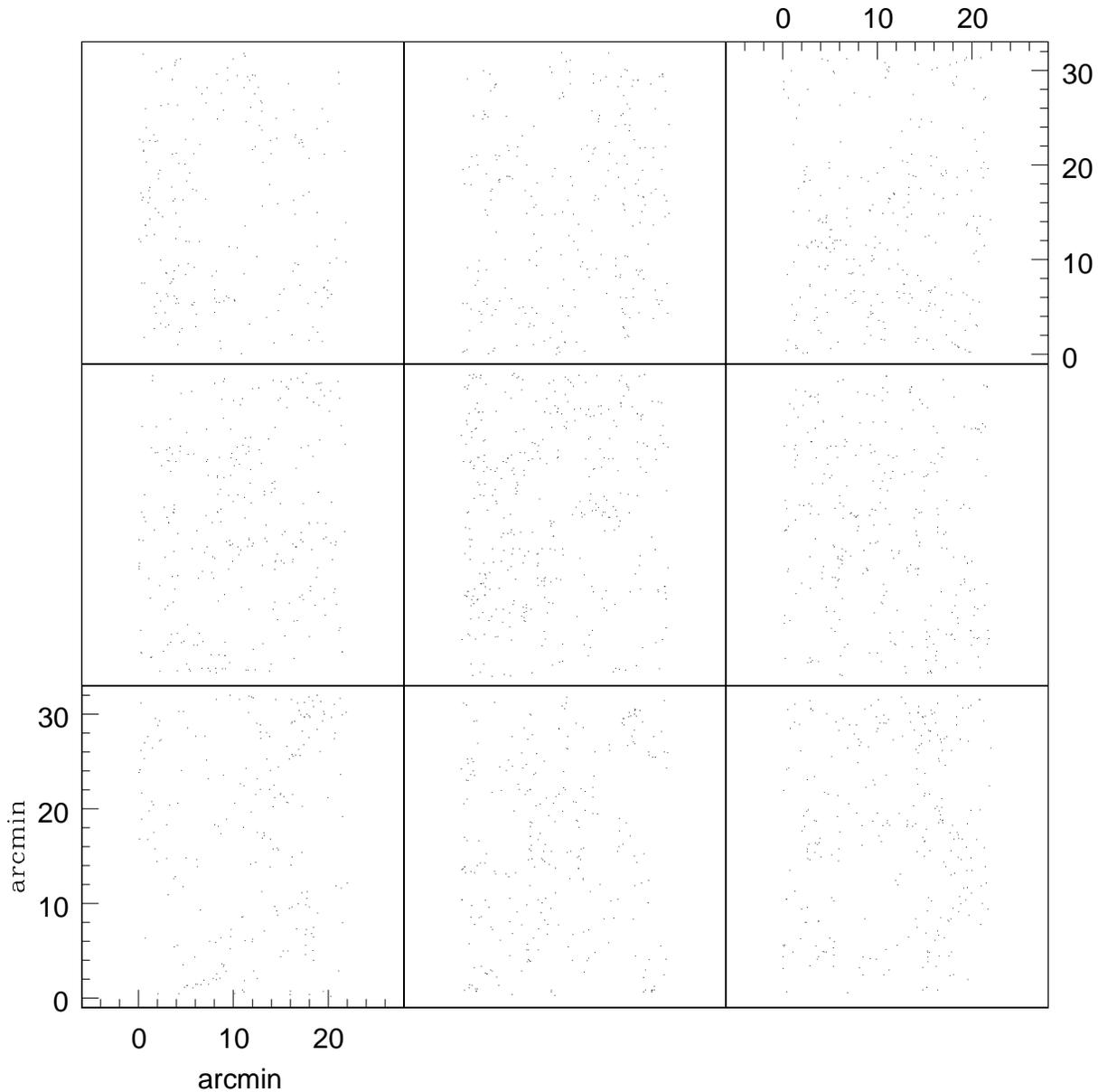}}
\caption{\footnotesize Examples of simulated realizations of our field
of $22\times 32$ arcmin populated with a population with fixed 3D
clustering $r_0=12$ \h1 Mpc. For reference, the top-right panel show
the sky distribution of our  data.  }
\label{fig:tom1}
\end{figure*}
\begin{figure}[ht]
\resizebox{\hsize}{!}{\includegraphics{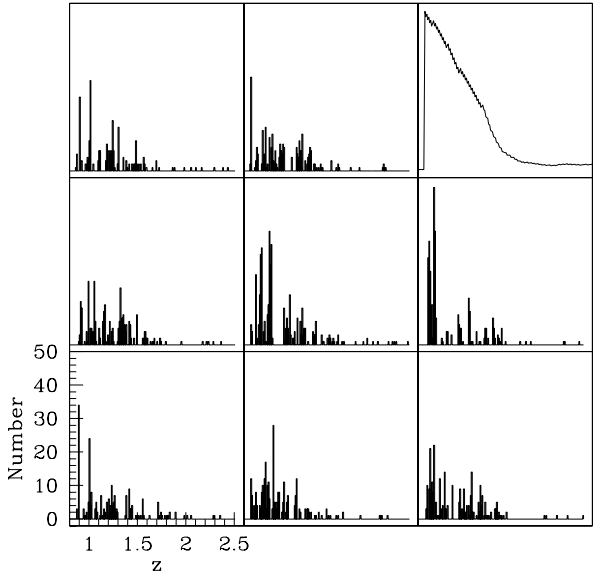}}
\caption{\footnotesize Examples of redshift distribution recovered from the
simulations. Each panel refers to the corresponding one in
Fig. \ref{fig:tom1}. In the top-right panel the adopted selection
function is shown.  }
\label{fig:tom2}
\end{figure}

\begin{figure}[ht]
\resizebox{\hsize}{!}{\includegraphics{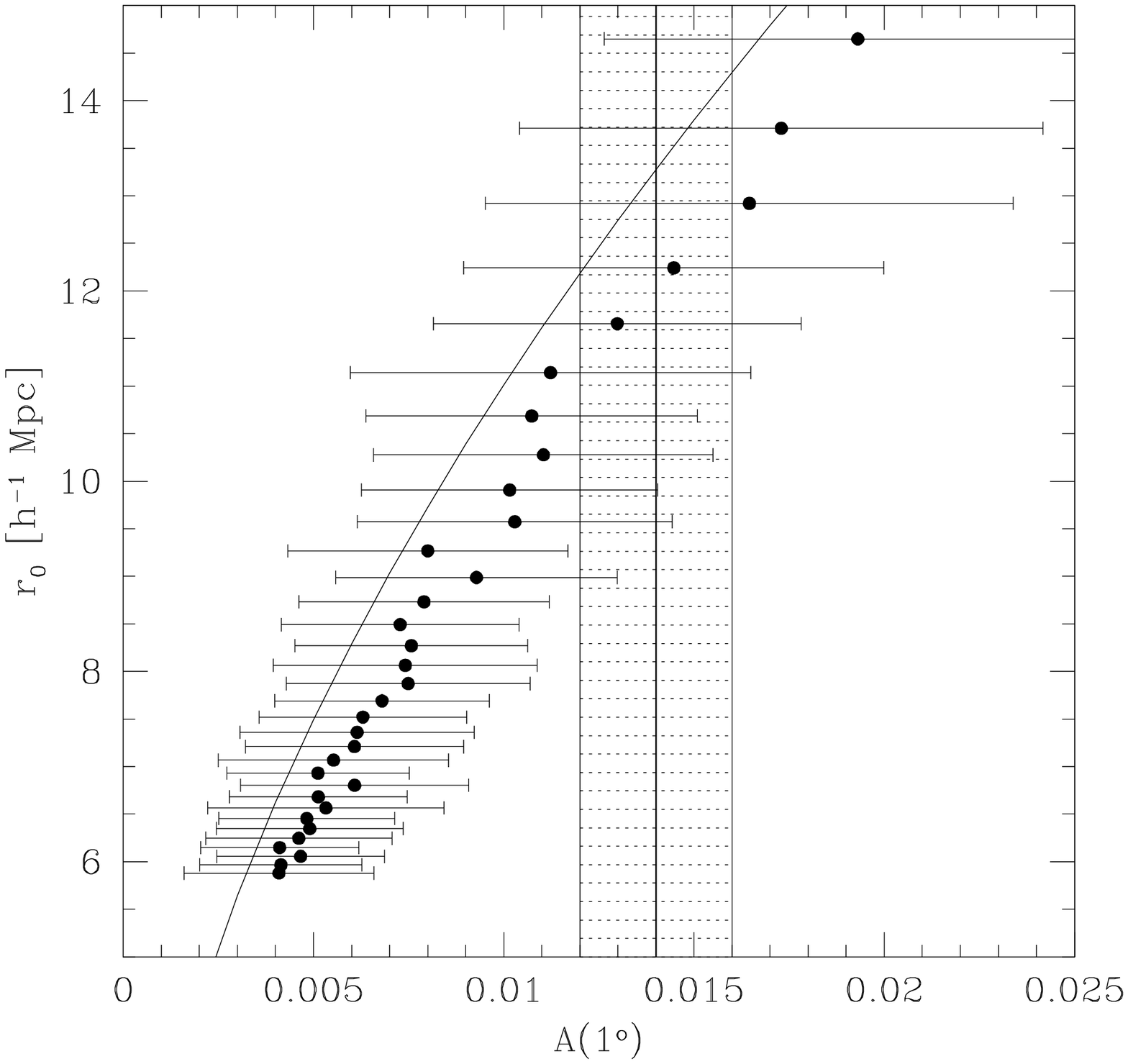}}
\caption{\footnotesize The data points show the mean angular clustering
amplitude recovered from the simulations, as a function of the input $r_0$. 
The error bars
correspond to the standard deviation of the distribution of the recovered 
amplitudes. For comparison, the
predictions of the Limber's equation are shown (curved line). The
shadowed area brackets our observed amplitude and its uncertainty
$A=0.014\pm0.002$.  }
\label{fig:Simul}
\end{figure}

The simulations were aimed at reproducing the observations for the EROs
with $Ks\leq18.8$ for which we estimate an angular
correlation of $A=0.014\pm0.002$, as this uses the
full 701 arcmin$^2$ area (the measurement at
$Ks=19.2$ has a better S/N but is limited to a smaller area of 
447.5 arcmin$^2$). 
Given that the analysis based on Limber's equation show that 
different $z_f$ and different cosmologies yield very similar results, we restricted our
simulations to the case of $dN/dz$ produced by the model with
$z_f=4$ and adopted a $\Lambda$-flat cosmology. The idea is to build up
a test case to better understand the behavior of projected clustering
since we do not expect this to depend significantly on the details of the
selection function.

We have produced 120 independent realizations of a field of view of
$22\times 32$ arcmin to match the data, for each of 33 values of
$r_0$ ranging from about 6 \h1 Mpc to 15 \h1 Mpc, and then we have
measured the angular clustering in the same way as for the data (basing 
on the Landy \& Szalay 1993 estimator, see
D2000). The simulations are populated in such a way to produce on
average 280 objects for each field to match the data for the EROs with $Ks\leq18.8$.  
In Fig. \ref{fig:tom1} and \ref{fig:tom2} we show some examples of simulated sky
projections together with their corresponding redshift
distributions, for a population with $r_0=12$ \h1 Mpc (that we show in section 
\ref{sect:r0eros} to be the best fitting value for EROs).

\begin{figure}[ht]
\resizebox{\hsize}{!}{\includegraphics{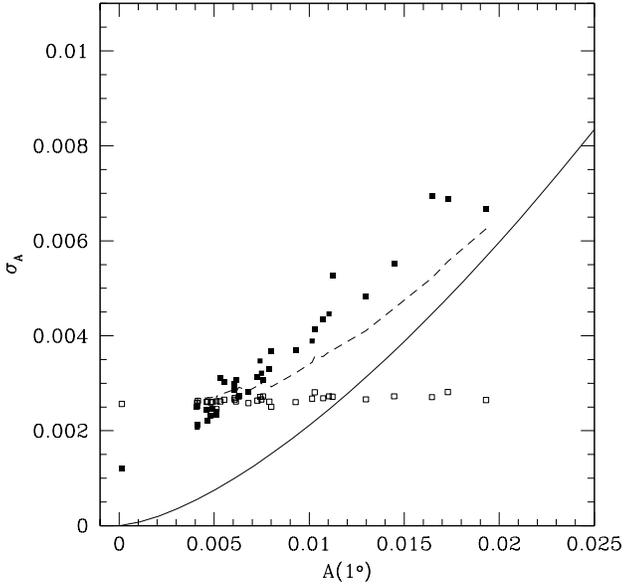}}
\caption{\footnotesize The filled squares show the standard deviation
versus the mean angular amplitude recovered from the simulations.  Empty
squares are the average statistical uncertainties in the $A$ measurements. The
solid line corresponds to the predictions of eq.  \ref{eq:dA}, while
dashed line is obtained by adding in quadrature both sources of
variance.  }
\label{fig:Error}
\end{figure}

The main results of the simulations are shown if Fig. \ref{fig:Simul}.
These simulations show two interesting findings. First, the presence of
a large dispersion in the measured values of the amplitude of the correlation
function, for any given assumed $r_0$ value. This in turns implies,
for any given measured amplitude, a large range for the statistically
acceptable values for $r_0$, well in excess of what obtained from the
Limber's equation (see previous Section). Secondly, for each $r_0$ value
the mean clustering amplitude recovered from the simulations is
systematically higher than what predicted by the Limber's equation.

\subsection{Intrinsic variance of the two-point correlation function}

The origin of both these effects (large dispersion and bias of the amplitude
of the correlation function) can be estimated from simple considerations.
The basic line of reasoning is the well known
fact that, because of clustering, the actual variance in the object
number counts is larger than the poissonian variance, and can be
written as (see D2000, eq. 8):
\begin{equation}
\sigma_n^2 = \overline{n}\ (1+\overline{n}AC)
\label{eq:sigma}
\end{equation}
where $\overline{n}$ is the mean expected number of objects {and $AC$ is the integral constraint
(Groth \& Peebles 1977):

\begin{equation}
AC = \frac{1}{\Omega^2} \int\int w(\theta) d\Omega_{\rm 1}d\Omega_{\rm 2}
\label{eq:AC}
\end{equation}
}

The expression for the variance in eq. \ref{eq:sigma} is the same which 
one would obtain if all the observed objects would belong to clumps with:
\begin{equation}
N_{\rm cl} = (AC)^{-1}
\label{eq:Ncl}
\end{equation}
(i.e., from eq. \ref{eq:AC}, the number of clumps is the inverse of the average of $w(\theta)$ over the observed field), 
and the number of objects per clump were a stochastic variable with
average $\overline{n}/N_{\rm cl}$. 
{From eq. \ref{eq:Ncl} it is expected that a variance in 
the number of clumps $N_{\rm cl}$ should result
in a variance in the clustering amplitude $A$. In the minimal hypothesis that
the clumps are distributed at random in the sky (i.e. neglecting the clustering between clumps) then:
\begin{equation}
\sigma_{N_{\rm cl}} \equiv \sqrt{N_{\rm cl}} \equiv \frac{\partial N_{\rm cl}}{\partial A} \sigma_A 
\end{equation}
from which it follows that the  relative dispersion on the amplitude of the 
correlation function caused by the sky fluctuation of $N_{\rm cl}$ would be}:
\begin{equation}
\frac{\sigma_A}{A} = \sqrt{AC}
\label{eq:dA}
\end{equation}

Thus, one should observe a real variation of
the correlation amplitude $A$ on the sky, even at fixed 3D clustering length.  
The observations of angular clustering for a population of fixed
$r_0$ should result in a distribution of values with a variance
decreasing with the area over which the measurements are carried out
(the factor $C$ is a decreasing function of the area, see
eq. 9 in D2000), and strongly increasing with the expected average
angular amplitude ($\sigma_A\propto A^{3/2}$).  

In any generic angular clustering measurement such
{\it intrinsic} variance, that depends on the survey geometry and the
clustering strength, has to be added to the statistical uncertainty in the
measurement of $A$ that is linked to the finite number of observed galaxies.  
In principle with very large areas (and/or weak
clustering) only the latter has a measurable effect, but in the case of
small fields, if the clustering itself is strong, the former may 
become the dominant source of uncertainty.

Our analytical derivation is supported by and can explain the results
of the simulations that we have carried out.
Fig. \ref{fig:Error} shows that only by adding both contributions, the
variance measured in the simulations can be recovered rather well, with some
underestimation ($\simlt10$\%) for large $A$ values.  Probably such
small underestimation is explained by other effects neglected here,
such as for instance the variance in the mean redshift of the clumps
in a given beam, as the angular clustering is increased for a 
lower mean redshift.
Incidentally, we note from Fig. \ref{fig:Error} that the 
statistical uncertainty in each single measurement
of $A$ is found to be rather constant as a function of $A$, thus
depending only on the number of observed objects. 
At the same time such statistical uncertainty seems to be overestimated by a factor of $\sim2$
(we recall that it follows from assuming 2 sigma poisson errors for the correlations
as suggested by bootstrap analysis, see the discussion about it in D2000),
given that $\lim_{A\rightarrow 0} \sigma_A$ is about half of the average statistical
uncertainty in the measurements of $A$ (see Fig. \ref{fig:Error}).

\begin{figure}[ht]
\resizebox{\hsize}{!}{\includegraphics{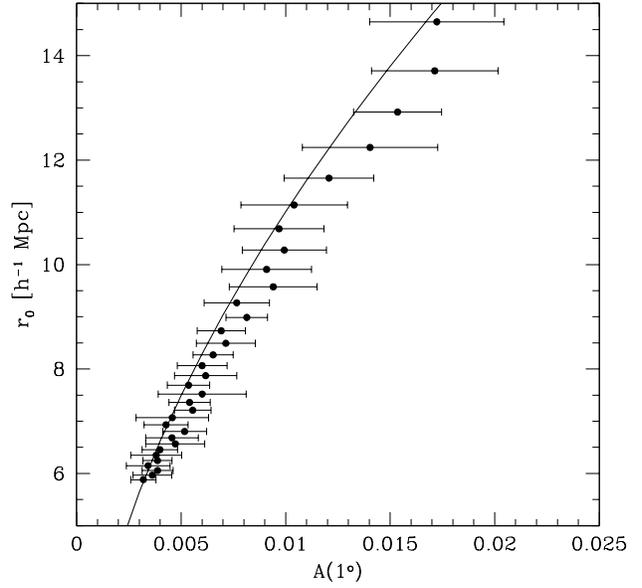}}
\caption{\footnotesize 
The same of Fig. \ref{fig:Simul}, but for a field of 7000 arcmin$^2$.
}
\label{fig:7000}
\end{figure}

The basic results of this analysis is that the variance in the angular
clustering can be much larger than the purely statistical variance, 
especially for small fields of view. 
A similar qualitative conclusion was empirically reached by Postman et al.
(1998) by splitting their large photometric survey into 250 smaller
subunits, finding a large scatter in the amplitudes recovered from the
smaller fields; a point taken up by McCracken et al. (2000) in the
analysis of a single very deep field of 50 arcmin$^2$. For the first
time here we quantify this effect and give a general analytical
prescription to predict its amplitude.  In the literature this
additional source of variance is usually not considered,
while deprojection analyses have been carried out even for surveys
covering tiny fields of view of only a few arcmin$^2$ of sky (e.g.
from the Hubble Deep Fields) which must severely underestimate the
true variance if the standard Limber's based inversion procedure is
applied.  Moreover, as a large variance is indeed expected for the
measurements of $A$, our findings may help to explain why so many
apparently discrepant clustering measurements (for similar observing
conditions) have been found in the literature (see e.g. Fynbo,
Freudling \& M\"oller 2000, McCracken et al. 2000).

\subsection{A possible bias in the Limber's equation based inversion}

The other interesting finding of our simulations is the possible
presence of a small (about 15\%) but significant bias, with respect to
Limber's equation predictions on the relation between $r_0$ and the measured
angular correlation amplitude (Fig. \ref{fig:Simul}).

\begin{figure}[ht]
\resizebox{\hsize}{!}{\includegraphics{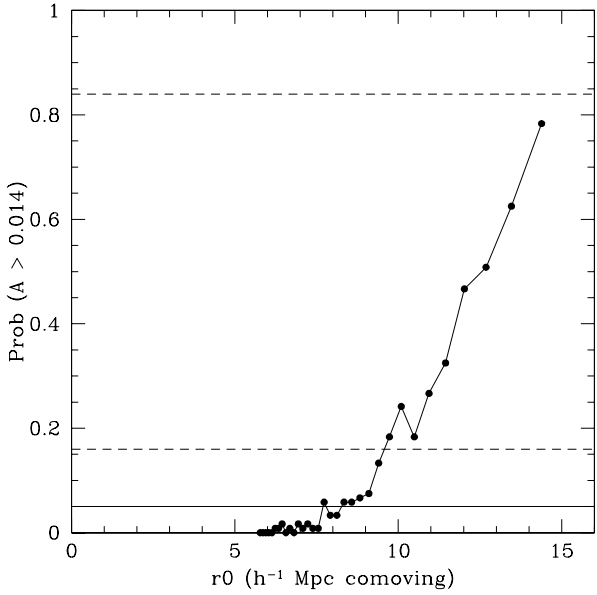}}
\caption{\footnotesize 
The probability to produce an ERO angular amplitude  $A>0.014$ as a function of the correlation
length $r_0$. The dashed lines show the $1 \sigma$ range for $r_0$.
}
\label{fig:Prob}
\end{figure}

Also this effect can be understood by thinking in terms
of objects coming in clumps, with $A$ and $N_{\rm cl}$ linked by
eq. \ref{eq:Ncl}. In fact, since in general $<\frac{1}{N_{\rm
cl}}>\neq\frac{1}{<N_{\rm cl}>}$, because of eq. \ref{eq:Ncl} we should expect deviations of $<A>$
from the Limber's equation predictions if $N_{\rm cl}$ is small or
if the distributions of $N_{\rm
cl}$ is significantly skewed. The inverse of $A$ (proportional to
$N_{\rm cl}$) should instead be less affected, and in
fact we find that the average of the {\it inverse} of $A$ is in better
agreement with Limber's equation predictions.  This effect is anyway small
compared to the huge variance and so for our sample it is not an
important correction. We have examined the possibility that this bias
is at least in part produced by our calculation procedures
generating some systematic effect, such for instance a bias in our
estimator (Hewett 1982; Kerscher et al. 2000) or an overcorrection for
the integral constraint. To rule out this possibility we built up
uncorrelated (random) samples with the same average number of objects
and measured their two point angular correlations. This results in an amplitude
of $A=1.4\ 10^{-4} \pm 1.2\ 10^{-3}$, thus showing that possible
systematic effects on the estimation of $A$ are much smaller than the
bias we find.  To verify further this point we carried out simulations
as in Sect. \ref{sec:simul}, over a larger area of 7000 arcmin$^2$
area (i.e. 2 square degrees, 10 times larger than our ERO field)
keeping the surface density of objects fixed at the observed
level. This exercise confirms the presence of some bias
although, as expected, at a lower level,
and still increasing with increasing
$r_0$ suggesting that it is a real effect (Fig. \ref{fig:7000}).
This larger simulation also allowed us to verify that the trend predicted by
eq. \ref{eq:dA} still holds correctly.

\subsection{Application of the simulations to EROs}
\label{sect:r0eros}

From Fig. \ref{fig:Simul} it can be seen that the ERO observed clustering
amplitude $A=0.014\pm0.002$ corresponds to  $r_0\sim 12 \pm3$ \h1 Mpc.  
This is considerably different from
the simple Limber's equation estimate of $r_0\sim13.2\pm0.8$ which we
infer underestimates the uncertainty in the clustering amplitude by a
factor of $\sim 3$.  
Thus, the cosmic variance is a substantial
source of uncertainty in the $r_0$ estimate for EROs,  
reducing the weight of the possible uncertainties in our selection function modeling. 
Using the spread in the $A$ measurements from the simulations
shown in fig. \ref{fig:Simul} we may place a lower limit to the
correlation length of $r_0\simgt8$ \h1 Mpc (see Fig. \ref{fig:Prob})
at the 95\% confidence level. While this estimate is derived only from
the measurement at $Ks=18.8$, we expect it to be consistent with what
we would have found from the analysis of the clustering at all $Ks$
levels. This is because within each single survey the number (and the
redshift) of the clumps is fixed in the real space, so that we expect
that the trend of the clustering amplitude as a function of the
apparent magnitude should basically reflect only the change of the
redshift distribution (see Fig. \ref{fig:AvsK}).

Our simulations have been carried out for convenience for the
$\Lambda$--flat universe. From the results of Sec. \ref{sec:Alimb} we
can say that in the open or $\Omega$--flat cases the $r_0$ estimate is
lower by a small amount, so that at the 95\% confidence level it
becomes $r_0\simgt7$ \h1 Mpc.

Fig. \ref{fig:AvsK} shows that the angular clustering measurement of
McCarthy et al. (2000) is consistent with our measurements,
even if slightly lower than our model predictions. 
Given the large variance expected, the
discrepancy is not significant. Nevertheless McCarthy et al. (2000)
estimate from their data $r_0\sim8$ \h1 comoving Mpc, significantly lower than our
preferred $r_0$ value. The main reason for this lower $r_0$ value is 
in their adoption of a relatively narrow Gaussian form of $dN/dz$,
centered at $z=1.2$ and with $\sigma_z=0.15$, motivated by their photometric
redshift estimates.  Such a strong confinement of EROs into a narrow
redshift range is not in agreement with our modeling of the
selection function shown in Fig. \ref{fig:dNdz}. Even if Fig. \ref{fig:tom2}
demonstrates that occasionally in small fields the observed $dN/dz$
could be spuriously narrow (because of the clustering), the inversion
process requires the use of the actual selection function, which we
expect to be much broader than the observed, clumpy $dN/dz$ of a 
given field. Assuming that the McCarthy et al. threshold of $I-H>3$ is consistent 
with $R-Ks>5$, and 
using our estimate of the ERO selection function at $Ks<19.5$, the McCarthy et al. 
angular measurement could be inverted to $r_0=10.8\pm2.2$, where the uncertainty is
derived from our own one
by keeping into account the scaling with the area and clustering amplitude, 
consistently with eq. \ref{eq:dA}.
This estimate apply to an effective redshift $z_{\rm eff}\sim1.2$ (Fig. \ref{fig:zbar}).

\section{Discussion}
\label{sec:evol}

\subsection{Comparison to the clustering of bright local ellipticals}

We now compare the large correlation length estimated for the
$z\simgt1$ ellipticals with that of their local counterparts, in order
to constrain the cosmic evolution of the clustering of massive
early-type galaxies.

The correlation length of a population of galaxies is known to depend on the
absolute luminosity selection threshold, and can also depend on the
scales over which the clustering is measured. Such quantities must be
properly estimated in order to compare the clustering of EROs to that
of local ellipticals.  For a passively evolving elliptical, the
apparent magnitude of $Ks=19.2$ corresponds to $L\sim 0.6L^{*}$ and
$L\sim1.3L^{*}$ at the redshift of 1 and 1.5, respectively, while for
$Ks=18$ we have $L\sim1.6L^{*}$ at $z=1$ and $L\sim4L^{*}$ at $z=1.5$
(accounting for the passive evolution of $L^{*}$).
Therefore our sample consists of galaxies with typical luminosity
$L\simgt L^{*}$.  The largest
effective separation probed by our angular clustering measurements is
$\theta\simeq 15^\prime$, corresponding to about 12 \h1 Mpc at $z\sim1$
($\Lambda$--flat universe).

As the clustering amplitude is expressed with $r_0^\gamma$, the
measurements of $r_0$ obtained with a $\gamma$ different from the
value adopted here must be rescaled to that value in order to produce
a meaningful comparison. Therefore, all the $r_0$ quoted below were
transformed with $\gamma=1.8$.

Our results can be compared with those obtained locally, for two different samples,
by Guzzo et al. (1997, the Perseus Pisces redshift survey) and
Willmer et al. (1998, the SSRS2 redshift survey). Both estimate the
clustering of bright early type galaxies with $M_B<-19.5+5logh$,
corresponding to $L\simgt L^{*}$. Guzzo et al. measure scales up to
about 10 \h1 Mpc and find $r_0=11.3\pm1.3$ \h1 Mpc, while Willmer et
al. measure up to about 20 \h1 Mpc and find $r_0=7.6\pm1.2$. The two
measurements are only consistent with each other at the 2$\sigma$
level, but it should be noted that the Perseus Pisces redshift survey
has a higher abundance of local clusters. It may be implied by Fig. 10
of Willmer et al. 1998 that they would obtain a larger amplitude if
limited to smaller separations.  For further constraints we note that local
radio galaxies, known to be hosted by bright ellipticals, have
$r_0=11\pm1.2$ \h1 Mpc (Peacock \& Nicholson 1991).  Therefore a
correlation length in the range $r_0\sim9$--11 \h1 Mpc can be regarded
as a reasonable estimate for the clustering amplitude of local $L\simgt L^{*}$
ellipticals.

Our estimate of $r_0=12\pm3$ \h1 Mpc then implies that the clustering
amplitude of bright ellipticals does not significantly decrease from
to $z\sim1-1.5$ to the present.  The {\it stable clustering} scenario
(i.e. $\epsilon=0$, if $r_0(z)=r_0(z=0)(1+z)^{(\gamma -3-\epsilon)/\gamma}$)
that is known to fit
many observations of clustering evolution to $z\sim1$ (e.g. Peebles 1980, Le Fevre
et al. 1996, Carlberg
et al. 1997), would predict  $r_0\sim6$ \h1 Mpc at our inferred effective
redshift of $z_{\rm eff}\sim1.1$ and hence it is not in good agreement with
our measurement of the ERO clustering. A negative value for $\epsilon$ is supported by our analysis.

\subsection{Comparison to theoretical predictions}

The evolution of the correlation function is popularly characterized as: \beq \xi
(z,r)=D^2(z)b^2(z)\xi_{mass} (0,r)
\label{eq:gcs}
\eeq where the purely linear growth $D(z)$ of density perturbations,
$\Delta \rho(z)/\rho(z)$ is separated from the bias evolution b(z).
In the linear case the bias evolution can be expressed with the
Tegmark \& Peebles (1998) prescriptions and well known expressions for
the linear growth factor can be assumed (Peebles 1980, Treyer \& Lahav
1996). In our case the linear assumptions are not likely to be
correct, as for EROs we are mapping angular comoving scales similar to the
inferred $r_0$, thus sampling a region with $\xi\sim1$, so that
this prescription can be considered as a rough benchmark picture 
in the absence of the complexities affecting small scale growth
(Kauffmann et al. 1999, Somerville et al. 2001).

The simplest and most clear model of bias evolution of ellipticals is
provided by the so called {\it galaxy conservation} scenario (Fry et
al. 1996, Tegmark \& Peebles 1998, Moscardini et al. 1998,
Magliocchetti et al. 1999, Lacy 2000), that holds if the galaxy
population is conserved over the cosmic time (i.e. no new elliptical
forms and no one disappears).  
This scenario implies 
the assumption that all the ellipticals were formed at high redshift and
simply follow the growth of perturbations without the additional
non-linear effects such as virial collapse and merging expected at
small scales (below $r_0$ of the mass auto-correlation function). It is
therefore relevant as a limiting case for comparison with our
observations.  Here the positive evolution of the bias which increases
with redshift is more than compensated by the decline of linear
growth, i.e. $D(z)$ wins and a net decline of the clustering amplitude
with increasing redshift is expected in the linear regime beyond
$r_0$ of the mass-autocorrelation function.

\begin{figure}[ht]
\resizebox{\hsize}{!}{\includegraphics{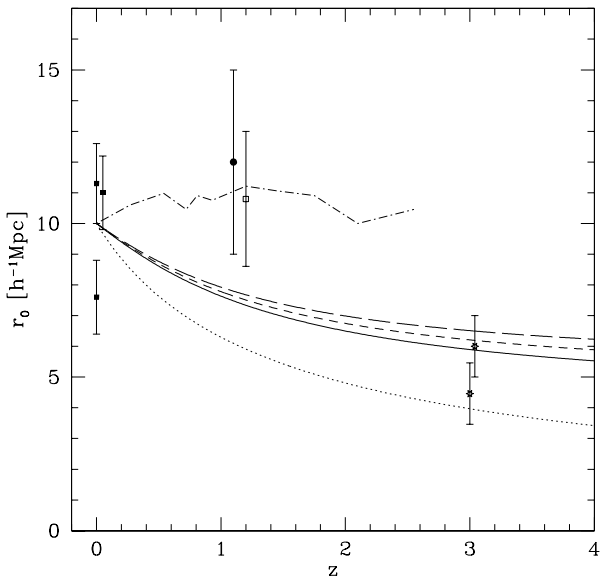}}
\caption{\footnotesize The $r_0$ measurements for local ellipticals (filled squares),
$z\sim1$ ellipticals (EROs; filled circle from D2000; the empty square is derived 
from our analysis applied  to the McCarthy et al. angular clustering measurement)
and LBGs (asterisks) are shown for the $\Lambda$-flat
universe (see the text for details).  The dot-dashed line shows the
Kauffmann et al. (1999) $\Lambda$CDM predictions for the clustering of ellipticals.
The case of stable clustering ($\epsilon=0$) is indicated
by the dotted line. The other three curves show the predictions for
the {\it galaxy conservation} scenario, differing in the assumed
degree of correlation between galaxy and dark matter, assumed from top
to bottom to be 0.9-0.95-1 at $z=0$ (see Tegmark and Peebles 1998, for more
details on this parameter).  All the models were normalized to
$r_0(0)=10$ \h1 Mpc.}
\label{fig:xiev}
\end{figure}

In Fig. \ref{fig:xiev} ($\Lambda$-flat case) we see that normalizing
the linear predictions to the $r_0$ and bias of local ellipticals (such
bias was computed assuming $r_0=10$ \h1 Mpc and the normalization
$\sigma_8^{mass}=0.9$, from Eke Cole \& Frenk 1996), the predicted
trend is slightly decreasing in comoving units, reaching values around
$r_0\sim7$--8 \h1 Mpc at $z=1.1$. Given the large uncertainties, this
scenario cannot be rejected with great confidence, but it is
disfavored by our findings.

If the {\it galaxy conservation} model predictions, both for the bias
and for $r_0$, are extended to higher $z$, 
they do intercept the measurements for the LBGs clustering by
Giavalisco et al. (1998) and Adelberger et al. (1998) (see
Fig. \ref{fig:xiev} for the $\Lambda$-flat case).  This kind of
argument had led to previous suggestions that LBG's could evolve into
present-day bright ellipticals (e.g. Adelberger 2000).  An
important confirmation of this picture would be to find that distant
(i.e. $z\sim 1$) ellipticals have intermediate clustering strength between the local
ones and the LBGs.  Our findings, taken at face value, disfavor such an interpretation as
they seem to suggest that the clustering of ellipticals is not
decreasing enough from $z\sim0$ to $z\sim1$ 
to reproduce the clustering at $z=3$, at least for the LBGs with 
absolute luminosity similar to those in the Giavalisco et al. (1998) 
and Adelberger et al. (1998) samples.

Alternative more complex models for the clustering evolution have been
constructed to deal with nonlinear effects and may be termed the {\it
galaxy merging} scenario for bias evolution (e.g. Mo \& White 1996,
Moscardini et al. 1998), in which the bias grows with $z$ with some
law of the kind $\Delta b\propto (1+z)^{1.8}$, which is stronger than
the growth of perturbations, with the net effect that $r_0$ increase
with $z$. Such models are in better agreement with the ERO clustering,
especially if the large favored $r_0\sim12$ \h1 Mpc will be confirmed.
Similar predictions for the clustering evolution of the ellipticals in
the $\Lambda$CDM semianalytic model of Kauffmann et al. (1999) and
Somerville et al. (2001) are also consistent with the trend inferred
here (see Fig. \ref{fig:xiev}) and include further refinements such as
luminosity evolution.

A difficulty one might expect of strong bias evolution models is that
they may contradict the observational evidence of the lack of evolution of the
comoving density of ellipticals, which appears not to decrease significantly
to $z=1$ and beyond (see Daddi Cimatti \& Renzini 2000). However, depending
on the details of the semi-analytical approach, it is possible to
accommodate substantial bias evolution of the dark matter without
perturbing either the apparent space density or clustering amplitude
of elliptical galaxies for the $\Lambda$CDM model (Kauffmann et
al. 1999; Somerville et al. 2001, see also Bullock et al. 2001) and the main difference
between these models and the conservation scenario would seem to
be the inclusion of a suitable prescription for merging.
As discussed in Daddi Cimatti \& Renzini (2000), the 
small amount of density evolution required by the
$\Lambda$CDM models could be consistent
with the ERO number counts once it is required that merging does not produce significant star 
formation (that would make the objects bluer than our color cuts), i.e. {\it red merging} is
required. This indeed is claimed to have been observed in clusters (van Dokkum et al. 2001).

\section{Summary and conclusions}
\label{sec:conclu}

The main results presented in this paper are:\\
$\bullet$ The real space correlation length of EROs (with $R-Ks>5$)
is much larger that the correlation of
$z\sim1$ star-forming galaxies for any reasonably large selection function, 
strengthening the previous suggestions that most EROs at $Ks\sim19$ are early-type galaxies.\\
$\bullet$ Assuming that EROs are predominantly early-type galaxies and hence that their
selection function is reasonably described by passive evolution, then the spatial
correlation length we obtain is rather large, not less than 7--8 \h1 
Mpc, with the most probable estimate of $12\pm3$ \h1 Mpc, applying to an effective redshift 
of $z\sim1.2$.\\
$\bullet$ At face value this implies no significant evolution of clustering of this
population relative to the present day when measured in {\it comoving
coordinates}, and a strong bias increase from $z=0$ to 1.\\
$\bullet$ Realistic simulations were used to constrain $r_0$ from the observed 
angular clustering of EROs.
Two main results follow from the simulations that can be
of general interest for the analysis of the angular
clustering: (1) the amplitude $A$ of the angular two-point correlation function
fluctuates on the sky according to $\sigma_A/A=\sqrt{AC}$, and (2) a possible systematic 
overestimate of $r_0$ could follow from the inversion of the angular clustering measurements
based on Limber equation. Both effects are strongly enhanced in the limits of strong angular
clustering and/or small fields of view.

Taken at face value, our result on the ERO correlation 
length challenges the simple conservation models of clustering
growth for massive haloes, but it may be reconciled with more complex
schemes for the bias which incorporate more parameters to describe
non-linear effects such as merging, so that although evolution of the
underlying mass function strongly declines with redshift, the observer
will, it is claimed, find the opposite of the expected behavior,
namely an increase in the observed correlation length of early-type
galaxies with redshift and no corresponding strong decline in their number
density with increasing redshift (Kauffmann et al. 1999, Somerville et
al. 2001). 
We rule out a high degree of density evolution of early type galaxies
to $z\sim1$. This is inconsistent with the ERO clustering because
strong density evolution would significantly narrow the width of the selection function by
removing the high-z tail resulting in a reduction of the inferred
estimate of $r_0$. This is counter to the strongly increasing
correlation amplitude that would be expected with increasing redshift for
such a highly biased model. A modest reduction in density at $z>1$
can be accommodated given the present uncertainties in lookback time
and star formation which fold into the construction of the selection
function. Indeed some change in density through merging is suggested
by our results when we combine the constraints on
both density and correlation length evolution.

Before discarding the {\it galaxy conservation} scenario for the
clustering evolution of early type galaxies some possible concerns
should anyway be carefully considered, that could make the measurement of
clustering spuriously high. Firstly if somehow the volume sampled by
EROs is over-abundant in rich clusters with respect to the local
samples, this would increase $r_0$.  In D2000 we discuss this point,
suggesting that it is unlikely and our simulations here support
this. Secondly if the EROs are somehow confined to a narrower redshift
range than expected on the basis of passive evolution, then the
estimate of $r_0$ should be lowered (McCarthy et al. 2000).
This would also have the effect
of increasing the comoving density, meaning in turn positive density
evolution which would be hard to imagine. Finally, we have evaluated
the cosmic variance with our modeling of the true {\it external} error bars
showing that we cannot rule out that both our result and that claimed by
McCarthy et al. (2000) are consistent with the {\it galaxy conservation} scenario, 
representing a $\approx 1.5\sigma$ high deviation from a true lower correlation length.
On the other hand, anyway, if a significant fraction of dusty objects is present among EROs
this would probably imply that the correlation length of the genuine early-type fraction 
could be larger than our estimate.

Much larger areas have to be observed to reduce the error on $r_0$. Our
simulations have shown that the variance of $A$ from spikes is a
slowly decreasing function of the area, while the statistical uncertainty in each
single $A$ measurement seems to decrease faster, with the square root
of the number of the objects, suggesting that the best strategy to get
rid of the eq. \ref{eq:dA} variance (and to increase thus the
precision on the estimate of $r_0$) is to observe many independent and
relatively large fields.  
At the same time the redshifts of complete samples of EROs should
be obtained to constrain their selection function. From Fig. \ref{fig:tom2}
we estimate that to observationally establish a detailed ERO selection function
will reasonably require thousands ERO redshifts to overcome the problems linked to the
existence of clumps.

\begin{acknowledgements}
We would like to thank the referee, Eelco van Kampen, Martin K\"ummel 
and Lucia Pozzetti for useful comments.

G.Z. acknowledges partial support by ASI (contracts ASI-ARS-99-15 and
ASI I/R/35/00) and MURST (Cofin 99).
\end{acknowledgements}


\begin{thebibliography}{}

\bibitem[1998]{adel}{Adelberger K., Steidel C., Giavalisco M., et al., 1998, ApJ 505, 18}

\bibitem[2000]{ad2000}{Adelberger K., 2000, Clustering at High Redshift, ASP Conference Series, Vol.
200. Edited by A. Mazure, O. Le Fevre, and V. Le Brun (astro-ph/9912153)}

\bibitem[2000]{andreani}{Andreani P., Cimatti A., Loinard L., R\"ottgering H., 2000, A\&A 354, L1}

\bibitem[1996]{bcf}{Baugh C.M., Cole S., Frenk C.S., 1996, MNRAS 283, 1361}

\bibitem[1990]{tom}{Broadhurst T., Ellis R., Koo D., \& Szalay A., 1990, Nature 343, 726}

\bibitem[2000]{brbow}{Broadhurst T., Bouwens R., 2000, ApJ 530, 53}

\bibitem[1993]{bc}{Bruzual G., Charlot S., 1993, ApJ 405, 538}

\bibitem[2001]{bull}{Bullock J., Dekel A., Kolatt T., et al., 2001, ApJ in press (astro-ph/0005325)}

\bibitem[1997]{carl}{Carlberg R., Cowie L., Songaila A. \& Hu E., 1997, ApJ 484, 538}

\bibitem[2001]{cims}{Cimatti A., 2001, to appear in the proceedings of the Deep Fields
Workshop, Garching (astro-ph/0012057)}

\bibitem[1998]{cmn}{Cimatti A., Andreani P., R\"{o}ttgering H., Tilanus R., 1998, Nature 392,
895}

\bibitem[1999]{cimanew}{Cimatti A., Daddi E., S. di Serego Alighieri, et al., 1999, A\&A, 352, L45}

\bibitem[1999]{cohen99}{Cohen. J, Hogg D., Blandford R., et al., 1999, ApJ 512, 30}

\bibitem[2000]{cohen2000}{Cohen. J, Hogg D., Blandford R., et al., 2000, ApJ 538, 29}

\bibitem[2000]{daddi2}{Daddi E., Cimatti A. \& Renzini A., 2000, A\&A 362, L45}

\bibitem[2000]{daddi}{Daddi E., Cimatti A., Pozzetti L., et al., 2000, A\&A 361, 535 (D2000)}

\bibitem[1976]{dav}{Davis M. \& Geller M., 1976, ApJ 208, 13}

\bibitem[1999]{dey}{Dey A., Graham J.R., Ivison R.J., et al., 1999, ApJ 519, 610}

\bibitem[1981]{dress}{Dressler A., 1980, ApJ 236, 351}

\bibitem[1991]{efs}{Efstathiou A., Bernstein G., Tyson A., et al., 1991, ApJ 380, L47}

\bibitem[1996]{ecf}{Eke V., Cole S. \& Frenk C., 1996, MNRAS 282, 263}

\bibitem[1998]{eise}{Eisenhardt P., Elston R., Stanford S.A., et al., proceedings of 
the X$^{\rm th}$ Rencontres de Blois (1998) on "The Birth of Galaxies", ed. B. Guiderdoni et al.
(astro-ph/0002468)}

\bibitem[1998]{franceschini}{Franceschini A., Silva L., Fasano G., et al., 1998, ApJ 506, 600}

\bibitem[2000]{fynbo}{Fynbo J., Freudling W. \& M\"oller P., 2000, A\&A 355, 37}

\bibitem[1996]{fry}{Fry J., 1996, ApJ 461, L65}

\bibitem[2000]{gear}{Gear W., Lilly S., Stevens J., et al., 2000, MNRAS 316, L51}

\bibitem[1998]{giava}{Giavalisco M., Steidel C., Adelberger K., et al., 1998, ApJ 503, 543}

\bibitem[1997]{guzzo}{Guzzo L., Strauss M., Fisher K., et al., 1997, ApJ 489, 37}

\bibitem[1986]{gio}{Giovanelli R., Haynes M. \& Chincarini G., 1986, ApJ 300, 77}

\bibitem[1977]{gp}{Groth E.J., Peebles P.J.E., 1977, ApJ 217, 38}

\bibitem[1982]{hewe}{Hewett P., 1982, MNRAS 201, 867}

\bibitem[2000]{hogg}{Hogg D., Cohen J. \& Blandford R., 2000, ApJ 545, 32}

\bibitem[1996]{kau96}{Kauffmann G., 1996, MNRAS, 281, 487}

\bibitem[1999]{k99}{Kauffmann G., Colberg J.M., Diaferio A. \& White S.D.M., 1999, MNRAS 307, 529}

\bibitem[2000]{kers}{Kerscher M., Szapudi I. \& Szalay A., ApJ 535, L13}

\bibitem[2000]{im}{Im M., Simard L., Faber S., et al., 2000, ApJ in press (astro-ph/0011092)}

\bibitem[2000]{lacy}{Lacy M., 2000, ApJ 536, L1}

\bibitem[1993]{ls}{Landy S.D., Szalay A.S., 1993, ApJ 412, 64}

\bibitem[1996]{lf}{Le Fevre O., Hudon L., Lilly S., et al., 1996, ApJ 461, 534}

\bibitem[1995]{love}{Loveday J., Maddox S.J., Efstathiou G., \& Peterson B.A., 1995, ApJ 442, 457}

\bibitem[2000]{liu}{Liu M.C., Dey A., Graham J.R., et al. 2000, AJ, 119, 2556}

\bibitem[1999]{mag1}{Magliocchetti M., Maddox S., Lahav O. \& Wall J., 1999, MNRAS 306, 943}

\bibitem[1994]{mkze}{Marzke R.O., Geller M.J., Huchra J.P., et al, 1994, AJ 108, 437}

\bibitem[2001]{mcca}{McCarthy P., et al., 2000, AAS 197, 6502  (see also astro-ph/0011499)}

\bibitem[2000]{mccr}{McCracken H., Shanks T., Metcalfe N., et al., 2000, MNRAS 318, 913}

\bibitem[1996]{mw}{Mo H. \& White S.D.M., 1996, MNRAS 282, 347}

\bibitem[2000]{morio}{Moriondo G., Cimatti A. \& Daddi E., 2000, A\&A 364, 26}

\bibitem[1998]{mosca}{Moscardini L., Coles P., Lucchin F. \& Matarrese S., 1998, MNRAS 299, 95}

\bibitem[1991]{pn}{Peacock J. \& Nicholson D., 1991, MNRAS 253, 307}

\bibitem[1980]{peebles}{Peebles P.J.E., 1980, The Large-Scale Structure of the Universe, Princeton University Press}

\bibitem[2000]{phill}{Phillipps S., Driver S., Couch W., et al., 2000, MNRAS 319, 807}

\bibitem[1998]{postm}{Postman M., Lauer T., Szapudi I., Oegerle W., 1998, ApJ 506, 33}

\bibitem[1999]{roche99}{Roche N., Eales S., 1999, MNRAS 307, 703}

\bibitem[1992]{sau}{Saunders W., Rowan-Robinson M. \& Lawrence A., 1992, MNRAS 258, 134}

\bibitem[1999]{shade}{Schade D., Lilly S.J., Crampton D., et al., 1999, ApJ 525, 31}

\bibitem[2000]{scode}{Scodeggio M., Silva D., 2000, A\&A 359, 953}

\bibitem[1999]{smail}{Smail I., Ivison R.J., Kneib J.P., et al., 1999, MNRAS 308, 1061}

\bibitem[1999]{soifer}{Soifer B.T., Matthews K., Neugebauer G., et al., 1999, AJ 118, 2065}

\bibitem[2001]{somer}{Somerville R., Lemson G., Sigad Y., et al., 2001, MNRAS 320, 289}

\bibitem[1977]{sp1}{Soneira S., Peebles P.J.E., 1977, ApJ 211, 1}

\bibitem[1978]{sp2}{Soneira S., Peebles P.J.E., 1978, AJ 83, 845}

\bibitem[1999]{spinrad}{Spinrad H., Dey A., Stern D., et al., 1997, ApJ 484, 581}

\bibitem[2000]{sttr}{Stiavelli M. \& Treu T., 2000, proceedings of the conference "Galaxy Disks and Disk Galaxies", ASP conference series, Funes and Corsini eds (astro-ph/0010100)}

\bibitem[1998]{tp}{Tegmark M. \& Peebles P.J.E., 1998, ApJ 500, L79}

\bibitem[1997]{ty}{Totani T., Yoshii J., 1997, ApJ 501, L177}

\bibitem[1996]{tl}{Treyer M. \& Lahav O., 1996, MNRAS 280, 469}

\bibitem[2001]{vdok}{van Dokkum P., Franx M., Fabricant D., et al., 2001, ApJ 541, 95}

\bibitem[1998]{will}{Willmer C., da Costa L. \& Pellegrini P., 1998, AJ 115, 869}

\bibitem[2001]{yosh}{Yoshida N., Colberg J., White S.D.M., et al., 2001, submitted to MNRAS
(astro-ph/0011212)}


\end{thebibliography}
\end{document}